\documentclass[12pt, letterpaper]{extarticle}

\usepackage{subcaption}
\usepackage{fullpage}
\usepackage[switch]{lineno}
\usepackage{amsmath}
\usepackage{amssymb}
\usepackage{rotating}
\usepackage{array}
\usepackage{mathtools}
\usepackage[ruled]{algorithm2e}
\usepackage{algorithmic}
\usepackage{bm}
\usepackage{breqn}
\usepackage{comment}
\usepackage{enumitem}
\usepackage{graphics}
\usepackage{graphicx}
\usepackage{latexsym}
\usepackage{mathrsfs}
\usepackage{morefloats}
\usepackage{nicefrac}
\usepackage{authblk}
\usepackage{pifont}
\usepackage{times}
\usepackage{xcolor}

\usepackage[hyphens]{url}
\usepackage{hyperref}
\hypersetup{colorlinks=false,breaklinks=true}

\usepackage{etoolbox}
\AtBeginEnvironment{quote}{\par\singlespacing\small}

\title{The illusion of neutrality in metric-based research evaluation}

\author[1+]{Fengyuan Liu}
\author[1+]{Hazem Ibrahim}
\author[1*]{Yasir Zaki}
\author[1*]{Talal Rahwan}

\affil[1]{\normalsize Computer Science, Science Division, New York University Abu Dhabi, UAE.}
\affil[+]{\footnotesize Joint first author}
\affil[*]{\footnotesize Corresponding authors. E-mail: \{yasir.zaki,talal.rahwan\}@nyu.edu}
\date{}

\begin{document} 

\maketitle 

\baselineskip22pt

\begin{abstract}
\noindent

Venue prestige and citation counts are two widely used, albeit imperfect, signals of research quality. When the two signals conflict, evaluators must decide how much weight to assign each. Yet, it remains unknown how researchers across disciplines trade off these two signals when evaluating research outcome. To fill this gap, we surveyed 869 researchers using paired choices between hypothetical departmental hiring rules that assigned different weights to venue prestige and citation counts, asking which would produce better science. Among 795 respondents whose choices were largely internally consistent, choices placed nearly equal aggregate weight on the two signals. This apparent balance concealed substantial individual heterogeneity: nearly two in five respondents occupied the most venue-heavy or citation-heavy intervals. Moreover, respondents on average chose more citation-heavy rules than they believed their departments used in hiring. By revealing the subjective judgments that arise when research indicators conflict, our findings reinforce calls for greater caution when quantitative indicators are used to evaluate research.
\end{abstract}

\clearpage

\section*{Introduction}

Research evaluation increasingly uses standardized indicators to promote consistency and transparency in judgments of scientific quality~\cite{Lim2025,hicks2015leiden}. Ideally, research outputs would be assessed by their quality~\cite{hicks2015leiden}, but in reality, researchers disagree over what quality entails~\cite{morales2021faculty}. Therefore, evaluators in practice rely on proxy indicators. Two of the most prominent such proxies are venue prestige and citation counts~\cite{Lim2025,pontika2022indicators,perneger2004relation}. Although publication in a prestigious venue typically carries a citation premium~\cite{traag2021inferring}, the two signals are not interchangeable and have become increasingly decoupled over time~\cite{lozano2012weakening}. Thus, standardized indicators do not eliminate judgment: their use requires a consequential choice about how different signals should be weighted.

Past research shows that scientists treat both venue-level and article-level signals as informative of scholarly value when the two are assessed independently~\cite{lemke2021conjoint}. When these two signals conflict, however, placing greater weight on one necessarily reduces the relative weight assigned to the other. The literature provides little direct evidence on this trade-off. Salandra et al.~examined a related trade-off among UK business and management academics, but focused on publication-outlet choices rather than evaluations of researchers and did not compare disciplines beyond business-school subfields~\cite{salandra2022academics}. Related surveys document discrepancies between researchers' own values and what they believe their peers or institutions reward~\cite{ross2024value,niles2020we,impey2025science}. However, because these studies assessed criteria independently, they did not examine whether researchers would weight venue prestige and article citations differently from their departments, how large any difference would be, or whether it would vary across disciplines.

To fill these gaps, we surveyed 869 researchers using paired choices between hypothetical departmental hiring rules that assigned different weights to venue prestige and citation counts. Respondents selected the rule they believed would produce stronger faculty and then identified the rule they believed most closely resembled their department's hiring practice; see Extended Methods for the survey materials. This design makes an important but usually implicit evaluative trade-off directly observable, allowing us to distinguish the aggregate pattern from the distribution of individual preferences and to compare those preferences with perceived departmental practice.

\begin{figure}[ht!]
    \centering
    \includegraphics[width=0.93\linewidth]{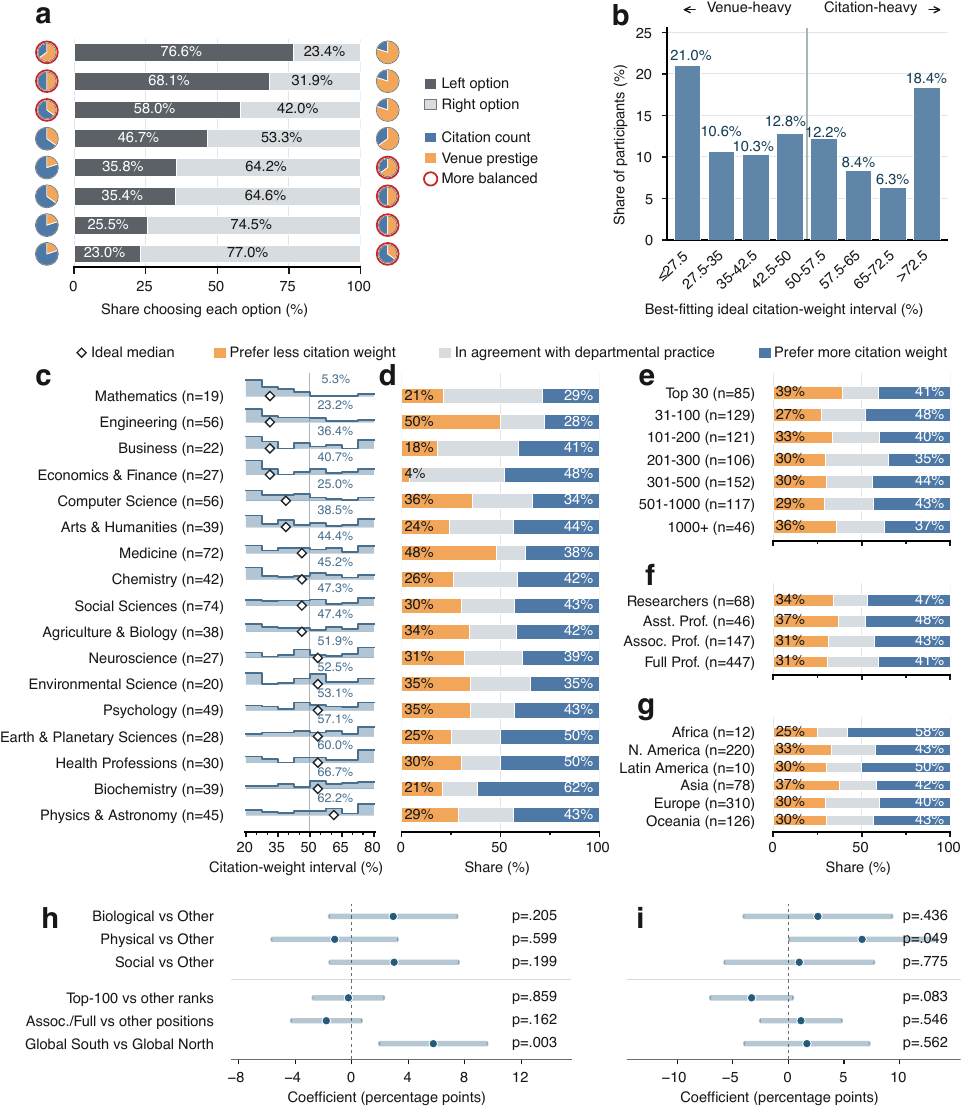}
\caption{
\textbf{Preferred evaluation rules and perceived departmental hiring practices.}
\textbf{a}, Percentage of the 869 respondents selecting each rule across eight unique paired comparisons. The ten survey choices included two pairs presented twice with their positions reversed; responses to each repeated pair were pooled. Pie icons show the weights assigned to citation counts (blue) and venue prestige (orange). Red outlines identify the rule closer to equal weighting when the two rules differed in this respect.
\textit{Continued on next page.}
}
    \label{fig}
\end{figure}
\begin{figure}[ht!]  
\caption*{
\textbf{b}, Distribution of best-fitting ideal citation-weight intervals among the 795 respondents whose ten choices contained at most one violation relative to their closest-fitting response profile. When several intervals fit equally well, the respondent's contribution was divided equally among them.
\textbf{c}, Distributions of best-fitting ideal citation-weight intervals across disciplines. Diamonds denote the median interval, the vertical line denotes equal weighting of citation counts and venue prestige, and the values at right report the percentage assigned to intervals strictly above 50\% citation weight.
\textbf{d}, Within-respondent comparisons between preferred ideal-weight intervals and perceived departmental-practice intervals, shown by discipline.
\textbf{e}, Same as (\textbf{d}), but by university-rank tier.
\textbf{f}, Same as (\textbf{d}), but by faculty rank.
\textbf{g}, Same as (\textbf{d}), but by institutional region. 
In (\textbf{e})--(\textbf{g}), orange indicates a preferred interval with less citation weight than perceived departmental practice, gray indicates that the preferred and perceived weights fall within the same interval, and blue indicates a preferred interval with more citation weight.
\textbf{h}, OLS coefficient estimates for absolute preference--practice mismatch. Positive coefficients indicate that the first-listed group exhibits a larger absolute mismatch than the corresponding reference group.
\textbf{i}, OLS coefficient estimates for signed preference--practice mismatch, defined as preferred citation weight minus perceived departmental citation weight. In (\textbf{h}) and (\textbf{i}), positive coefficients indicate that the first-listed group exhibits a more citation-oriented gap than the corresponding reference group. Points denote coefficient estimates, horizontal lines denote 95\% confidence intervals, and two-sided \(p\)-values are reported at right.
}
\end{figure}

\section*{Results}

Figure~\ref{fig}a reports the share of respondents selecting each rule across the eight paired comparisons. In each of the seven comparisons in which one rule was closer to an equal weighting of citation counts and venue prestige than the other, a majority selected that rule (58.0\%--77.0\%).

We next ask whether this aggregate balance reflects broad agreement or conceals substantial heterogeneity in respondents' preferences. To this end, we use each respondent's choices across comparisons to infer their preferred weighting interval---the range of allocations most consistent with their judgments about which hiring rules would produce stronger faculty. This inference assumes that each respondent has a stable preferred allocation of weight between the two signals and tends to select, in each comparison, the rule closer to that allocation. For example, a respondent whose ideal weight allocation is 30\%--70\% (citation counts and venue prestige, respectively) would be expected to select 20\%--80\% over 50\%--50\%, even though neither option exactly matches that ideal. By contrast, selecting 30\%--70\% over 50\%--50\% in one comparison but 70\%--30\% over 50\%--50\% in another would be inconsistent with any single ideal weighting. For each respondent, we therefore identify the ideal-weight interval that minimizes the number of violations across their choices (see Extended Methods). Of the 869 respondents, 603 (69.4\%) made all 10 choices without any violations, and 795 (91.5\%) made at most one. Thus, for the vast majority of respondents, the observed choices were highly consistent with a stable ideal weighting. Subsequent analyses focus on these 795 respondents.

Across these respondents, the mean inferred citation weight was 48.3\%, indicating a nearly equal aggregate weighting of the two signals. This aggregate balance, however, concealed substantial individual heterogeneity (Figure~\ref{fig}b). Only 25.0\% of respondents fell within the two intervals immediately surrounding equal weighting (42.5\%--57.5\% citation weight). By contrast, 21.0\% fell within the most venue-heavy interval ($\leq 27.5\%$ citation weight), and 18.4\% fell within the most citation-heavy interval ($>72.5\%$). Thus, 39.4\% of respondents occupied one of the two outermost intervals despite the nearly balanced aggregate estimate.
Preferred weight allocations varied across disciplines (Figure~\ref{fig}c). Among the 17 disciplines included in this comparison, the share of respondents preferring a citation weight above 50\% ranged from 5.3\% in Mathematics to 66.7\% in Biochemistry. Nevertheless, substantial heterogeneity remained within every discipline.

We then compared each respondent's preferred rule with the rule they believed most closely resembled their department's hiring practice. For 26.2\% of respondents, the preferred rule and perceived departmental practice fell within the same interval. By contrast, 42.1\% preferred a more citation-heavy rule than the one they attributed to their department, whereas 31.8\% preferred a more venue-heavy rule. On average, respondents' preferred rules assigned 3.9 percentage points more weight to citations than the rules they attributed to their departments (95\% CI: 2.3 to 5.5; paired-samples \(t\)-test, \(t(794)=4.82\), \(p<.001\)). Descriptively, the share preferring a more citation-heavy rule exceeded the share preferring a more venue-heavy rule in 13 of the 17 disciplines (Figure~\ref{fig}d) and in every university-rank tier (Figure~\ref{fig}e), faculty-rank group (Figure~\ref{fig}f), and geographic region (Figure~\ref{fig}g).

Adjusted OLS models indicated that respondents at Global South institutions had a larger absolute preference--practice mismatch than those at Global North institutions (+5.77 percentage points, \(p=.003\)), whereas the direction of the mismatch did not clearly differ between the two groups (+1.67 percentage points, \(p=.562\)). This suggests that respondents at Global South institutions differed more from their perceived departmental practice, but not in a systematically more citation- or venue-oriented direction. Apart from a nominal association between physical-science affiliation and the direction of the mismatch (\(p=.049\)), the remaining subgroup associations were not statistically significant. Given the number of comparisons, these subgroup results should be interpreted as exploratory.

\section*{Discussion}

Although evaluation committees are unlikely to rank candidates by applying a literal weighted average of citation counts and venue prestige, the stylized choices in our survey render a consequential but usually implicit trade-off between the two signals quantitatively observable~\cite{arabi2025most,steck2020journal}. At the aggregate level, respondents assigned nearly equal weight to the two signals, but this apparent balance concealed substantial heterogeneity in their preferred allocations. To our knowledge, this study provides the first cross-disciplinary quantitative estimate of how researchers trade off venue prestige and citation counts when evaluating researchers. We also find that respondents preferred, on average, more citation-heavy rules than those they attributed to their departments.

This trade-off reflects a deeper tension between journal-level and article-level indicators used in research evaluation. Consistent with Goodhart's law, a citation-dominated system risks turning citation counts from a measure into a target, encouraging strategic self-citation, coercive citation, and even purchased citations~\cite{fong2017authorship,ibrahim2025citation}. A prestige-dominated system, by contrast, risks perpetuating inequality by concentrating evaluative authority within a narrow gatekeeping class whose composition underrepresents large parts of the scientific community~\cite{liu2023non} and whose judgments may be influenced by professional connections and intellectual proximity~\cite{liu2025current,li2017expertise}.

Although balancing citation counts and venue prestige may mitigate the risks of relying exclusively on either, it does not resolve the broader limitations of indicator-based assessment. Standardized indicators can appear more objective than qualitative judgments because they yield comparable quantities. Yet determining how much weight to assign each remains a value judgment. The absence of a shared preferred allocation, even within disciplines, shows that quantitative evaluation does not eliminate human judgment. By revealing the subjective judgments that arise when seemingly objective research indicators conflict, our findings reinforce calls for greater caution when quantitative indicators are used to evaluate research~\cite{hicks2015leiden}.

\section*{Methods}

The survey was pre-registered on AsPredicted before data collection (\#256154; pre-registered 4 November 2025) and approved by the New York University Abu Dhabi Institutional Review Board (protocol HRPP-2025-248, exempt). Participation was voluntary and uncompensated. The survey was not piloted prior to deployment.

\section*{Data and Code Availability}
All survey data and code used to analyze this data are made publicly available in \url{https://github.com/Michael98Liu/citation-prestige-tradeoff}.

\bibliographystyle{naturemag}
\bibliography{sample}

\section*{Author Contributions}
T.R. and Y.Z. conceived the study; H.I., T.R., and Y.Z. designed the research; H.I. collected the data; H.I. and F.L analyzed the data; H.I., F.L., T.R., and Y.Z. wrote the manuscript.

\section*{Competing Interests}
The authors declare no competing interests.

\clearpage
\section*{Extended Methods}

\begingroup
\newcommand{\FHPRecTitle}[1]{%
  \par\noindent{\fontsize{18}{22}\selectfont\bfseries #1\par}\vspace{1em}%
}
\FHPRecTitle{Extended Methods Section 1: Survey recruitment, exclusions, and pre-registration}

The study was pre-registered on AsPredicted before data collection (\#256154; pre-registered 4 November 2025; \url{https://aspredicted.org/iv78m6.pdf}) and approved by the New York University Abu Dhabi Institutional Review Board (protocol HRPP-2025-248, exempt). Participation was voluntary and uncompensated, and no identifiable information was collected.

We recruited research-active scientists affiliated with universities ranked among the top 500 in the 2026 QS World University Rankings. Invitations were emailed to 92,925 researchers, of whom 1,166 completed the survey, corresponding to a completion rate of approximately 1.3\%. This rate was close to the approximately 1\% anticipated in the pre-registered sampling plan. Because participation was voluntary and uncompensated, researchers with stronger views about research evaluation may have been more likely to participate. Our estimates should therefore be interpreted as describing this self-selected sample rather than as population estimates for all invited researchers.

Following the pre-registered exclusion criteria, we excluded respondents who (i) incorrectly answered a comprehension check asking which signal received greater weight under a rule assigning 70\% to venue prestige and 30\% to citations, (ii) did not complete all paired comparisons, or (iii) completed the survey in less than one-third of the median completion time. Out of all 1,166 who consented and completed all survey questions, 57 were filtered out due to completing the survey in less than one-third of the median completion time, resulting in 1,109 remaining participants, we further filter out 240 respondents who did not correctly answer the comprehension check question. The resulting analytic sample comprised 869 respondents (74.5\% of the 1,166 completers).

\paragraph{Deviations from the pre-registration.} The pre-registration specified that ideal weights would be inferred from the paired choices using a mixed or conditional logit model. During analysis, inspection of the respondent-level choice patterns indicated that most respondents' choices were highly consistent with a stable ideal weighting. We therefore inferred each respondent's ideal-weight interval using the minimum-violations procedure described in Extended Methods Section~3. This procedure represents each respondent's choice pattern directly and makes inconsistencies observable without imposing a logistic functional form on choice probabilities. Hence we went with this procedure instead of the preregistered approach.

\endgroup

\clearpage
\begingroup
\setlength{\parindent}{0pt}
\setlength{\parskip}{0.6em}
\setlength{\fboxsep}{6pt}
\renewcommand{\labelitemi}{$\circ$}

\newcommand{\FHPLineBreak}{\leavevmode\newline}
\newcommand{\FHPSpaces}[1]{\hspace*{#1\fontdimen2\font}}
\newcommand{\FHPTitle}[1]{%
  \par\noindent{\fontsize{18}{22}\selectfont\bfseries #1\par}\vspace{1em}%
}
\newcommand{\FHPBlockLabel}[1]{%
  \par\noindent{\bfseries #1}\par%
}
\newcommand{\FHPBlockSeparator}{%
  \par\noindent\rule{\linewidth}{0.8pt}\par%
}
\newcommand{\FHPQuestionSeparator}{%
  \par\noindent
  \makebox[\linewidth]{\leaders\hbox{\kern0.2em--\kern0.2em}\hfill\kern0pt}\par%
}
\newcommand{\FHPSkipLogic}[1]{%
  \par\noindent\fbox{%
    \parbox{\dimexpr\linewidth-2\fboxsep-2\fboxrule\relax}{\itshape\small #1}}\par%
}

\FHPTitle{Extended Methods Section~2: Survey Vignette}
\FHPBlockSeparator
\FHPBlockLabel{Start of Block: consent\_block}
\par\noindent You are being asked to provide consent to participate in a research study. Participation is voluntary. You can say yes or no. If you say yes now you can still change your mind later.\FHPLineBreak{}
\textbf{Purpose of Research:}~This research is being conducted to better understand how academics across different fields perceive the prestige of scholarly journals, and what characteristics contribute to perceptions of prestige.\FHPLineBreak{}
\textbf{Procedures:}~You will be asked to complete a short online survey consisting of multiple-choice and open-ended questions about your field, which journals you consider prestigious, and the characteristics that make journals prestigious.\FHPLineBreak{}
\textbf{Duration:}~Participation will involve approximately about 7\textbf{~minutes of your time.}\FHPLineBreak{}
\textbf{Risks:}~We believe there are no known risks associated with this research study. Your participation in this research is voluntary, and you will not be penalized or lose benefits if you refuse to participate or decide to stop.\FHPLineBreak{}
\textbf{Benefits:}~There are no direct personal benefits to you from participating in this study. However, the results may help researchers and the academic community better understand how prestige is constructed across scholarly publishing, which could inform policy and practice.\FHPLineBreak{}
\textbf{Compensation:}~You will not receive financial compensation for participation in this study.\FHPLineBreak{}
\textbf{Privacy and Confidentiality:}~\textbf{No identifiable information (e.g., your name and email address) will be collected.} The researchers will keep all study records stored in a secure location. All electronic files (e.g., databases, spreadsheets, etc.) containing identifiable information will be password protected. Any computer hosting such files will also have password protection to prevent access by unauthorized users. Only the members of the research staff will have access to the passwords. Data that will be shared with others will be anonymized to help protect your identity. At the conclusion of this study, the researchers may publish their findings. Information will be presented in summary format and you will not be identified in any publications or presentations.\FHPLineBreak{}
You should also know that the NYUAD Institutional Review Board (IRB) may inspect study records as part of its auditing program, but these reviews will only focus on the researchers and not on your responses or involvement. The IRB is a group of people who review research studies to protect the rights and welfare or research participants. If you have further questions about this study or if you have a research-related problem, you may contact the principal investigators, Yasir Zaki or Talal Rahwan (nyuad.ai-and-society@nyu.edu). If you have any questions concerning your rights as a research participant, you may contact the New York University Abu Dhabi Institutional Review Board (IRB) at irbnyuad@nyu.edu.\FHPLineBreak{}
By clicking the button below, you acknowledge:\FHPLineBreak{}
1. Your participation in the survey is voluntary.\FHPLineBreak{}
2. You are 18 years of age or older.\FHPLineBreak{}
3. You are aware that you may withdraw from the study at any time for any reason.\par
\begin{itemize}
\item I consent\FHPSpaces{2}(1) 
\item I do not consent\FHPSpaces{2}(2) 
\end{itemize}
\FHPSkipLogic{Skip To: End of Survey If You are being asked to provide consent to participate in a research study. Participation is volun... = I do not consent}
\FHPBlockLabel{End of Block: consent\_block}
\FHPBlockSeparator
\FHPBlockLabel{Start of Block: question\_block}
\par\noindent area\_of\_research Which academic field best describes your \textbf{primary area of research}?\par
\begin{itemize}
\item Agricultural and Biological Sciences\FHPSpaces{2}(62) 
\item Arts and Humanities\FHPSpaces{2}(63) 
\item Biochemistry, Genetics and Molecular Biology\FHPSpaces{2}(64) 
\item Business, Management and Accounting\FHPSpaces{2}(65) 
\item Chemical Engineering\FHPSpaces{2}(66) 
\item Chemistry\FHPSpaces{2}(67) 
\item Computer Science\FHPSpaces{2}(68) 
\item Decision Sciences\FHPSpaces{2}(69) 
\item Dentistry\FHPSpaces{2}(70) 
\item Earth and Planetary Sciences\FHPSpaces{2}(71) 
\item Economics, Econometrics and Finance\FHPSpaces{2}(72) 
\item Energy\FHPSpaces{2}(73) 
\item Engineering\FHPSpaces{2}(74) 
\item Environmental Science\FHPSpaces{2}(75) 
\item Health Professions\FHPSpaces{2}(76) 
\item Immunology and Microbiology\FHPSpaces{2}(77) 
\item Materials Science\FHPSpaces{2}(78) 
\item Mathematics\FHPSpaces{2}(79) 
\item Medicine\FHPSpaces{2}(80) 
\item Neuroscience\FHPSpaces{2}(81) 
\item Nursing\FHPSpaces{2}(82) 
\item Pharmacology, Toxicology and Pharmaceutics\FHPSpaces{2}(83) 
\item Physics and Astronomy\FHPSpaces{2}(84) 
\item Psychology\FHPSpaces{2}(85) 
\item Social sciences\FHPSpaces{2}(86) 
\item Veterinary\FHPSpaces{2}(87) 
\item Other\FHPSpaces{2}(88) \_\_\_\_\_\_\_\_\_\_\_\_\_\_\_\_\_\_\_\_\_\_\_\_\_\_\_\_\_\_\_\_\_\_\_\_\_\_\_\_\_\_\_\_\_\_\_\_\_\_
\end{itemize}
\FHPQuestionSeparator
\par\noindent *\par
\par\noindent subfield What \textbf{subfield}~best describes your \textbf{primary area of research}?\par
\par\noindent \_\_\_\_\_\_\_\_\_\_\_\_\_\_\_\_\_\_\_\_\_\_\_\_\_\_\_\_\_\_\_\_\_\_\_\_\_\_\_\_\_\_\_\_\_\_\_\_\_\_\_\_\_\_\_\_\_\_\_\_\_\_\_\_\par
\FHPQuestionSeparator
\par\noindent position What is your current \textbf{academic position}?\par
\begin{itemize}
\item Graduate Student\FHPSpaces{2}(1) 
\item Postdoc\FHPSpaces{2}(2) 
\item Assistant Professor\FHPSpaces{2}(3) 
\item Associate Professor\FHPSpaces{2}(4) 
\item Full Professor\FHPSpaces{2}(5) 
\item Research Scientist\FHPSpaces{2}(6) 
\item Adjunct or Visiting Faculty\FHPSpaces{2}(7) 
\item Other (please specify)\FHPSpaces{2}(8) \_\_\_\_\_\_\_\_\_\_\_\_\_\_\_\_\_\_\_\_\_\_\_\_\_\_\_\_\_\_\_\_\_\_\_\_\_\_\_\_\_\_\_\_\_\_\_\_\_\_
\end{itemize}
\FHPQuestionSeparator
\par\noindent institution What university are you primarily \textbf{affiliated} with?\par
\par\noindent \_\_\_\_\_\_\_\_\_\_\_\_\_\_\_\_\_\_\_\_\_\_\_\_\_\_\_\_\_\_\_\_\_\_\_\_\_\_\_\_\_\_\_\_\_\_\_\_\_\_\_\_\_\_\_\_\_\_\_\_\_\_\_\_\par
\FHPBlockLabel{End of Block: question\_block}
\FHPBlockSeparator
\FHPBlockLabel{Start of Block: Consent and Comprehension}
\par\noindent Q15 \textbf{IMPORTANT (Please read carefully):}\FHPLineBreak{}
You will evaluate hypothetical faculty-hiring protocols. Each protocol specifies how much weight a department places on \textbf{citation counts} vs. \textbf{venue prestige} when evaluating candidates (the greater the weight, the greater the department's emphasis on this factor).\FHPLineBreak{}
Definitions:\FHPLineBreak{}
1.\textbf{ Citation counts} refers to the total number of citations the applicant has accumulated when applying to the department.\FHPLineBreak{}
2. \textbf{Venue prestige}~refers to the respect and admiration given to the venues (journals, conferences, book publishers, etc.) usually because of a reputation for high quality, success, or social influence.\FHPLineBreak{}
Percentages sum to 100. There are no right or wrong answers.\par
\FHPQuestionSeparator
\par\noindent Q16 If a department assigns 70\% weight to venue prestige and 30\% to citation counts, which factor is emphasized more?\par
\begin{itemize}
\item Citation counts\FHPSpaces{2}(1) 
\item Venue prestige\FHPSpaces{2}(2) 
\item They are equal\FHPSpaces{2}(3) 
\end{itemize}
\FHPSkipLogic{Skip To: End of Survey If If a department assigns 70\% weight to venue prestige and 30\% to citation counts, which factor is... != Venue prestige}
\FHPBlockLabel{End of Block: Consent and Comprehension}
\FHPBlockSeparator
\FHPBlockLabel{Start of Block: Choice Experiment 2080 3565}
\par\noindent 2080\_3565 Which department do you believe would produce stronger faculty (produce better science)?\FHPLineBreak{}
\textbf{Weights in Department A:}\FHPLineBreak{}
Citation counts: 20\%\FHPLineBreak{}
Venue prestige: 80\%\FHPLineBreak{}
\textbf{Weights in Department B:}\FHPLineBreak{}
Citation count: 35\%\FHPLineBreak{}
Venue prestige: 65\%\par
\begin{itemize}
\item Department A\FHPSpaces{2}(1) 
\item Department B\FHPSpaces{2}(2) 
\end{itemize}
\FHPBlockLabel{End of Block: Choice Experiment 2080 3565}
\FHPBlockSeparator
\FHPBlockLabel{Start of Block: Choice Experiment 2080 5050}
\par\noindent 2080\_5050 Which department do you believe would produce stronger faculty (produce better science)?\FHPLineBreak{}
\textbf{Weights in Department A:}\FHPLineBreak{}
Citation counts: 20\%\FHPLineBreak{}
Venue prestige: 80\%\FHPLineBreak{}
\textbf{Weights in Department B:}\FHPLineBreak{}
Citation count: 50\%\FHPLineBreak{}
Venue prestige: 50\%\par
\begin{itemize}
\item Department A\FHPSpaces{2}(1) 
\item Department B\FHPSpaces{2}(2) 
\end{itemize}
\FHPBlockLabel{End of Block: Choice Experiment 2080 5050}
\FHPBlockSeparator
\FHPBlockLabel{Start of Block: Choice Experiment 3565\_6535}
\par\noindent 3565\_6535 Which department do you believe would produce stronger faculty (produce better science)?\FHPLineBreak{}
\textbf{Weights in Department A:}\FHPLineBreak{}
Citation counts: 35\%\FHPLineBreak{}
Venue prestige: 65\%\FHPLineBreak{}
\textbf{Weights in Department B:}\FHPLineBreak{}
Citation count: 65\%\FHPLineBreak{}
Venue prestige: 35\%\par
\begin{itemize}
\item Department A\FHPSpaces{2}(1) 
\item Department B\FHPSpaces{2}(2) 
\end{itemize}
\FHPBlockLabel{End of Block: Choice Experiment 3565\_6535}
\FHPBlockSeparator
\FHPBlockLabel{Start of Block: Choice Experiment 3565\_8020}
\par\noindent 3565\_8020 Which department do you believe would produce stronger faculty (produce better science)?\FHPLineBreak{}
\textbf{Weights in Department A:}\FHPLineBreak{}
Citation counts: 35\%\FHPLineBreak{}
Venue prestige: 65\%\FHPLineBreak{}
\textbf{Weights in Department B:}\FHPLineBreak{}
Citation count: 80\%\FHPLineBreak{}
Venue prestige: 20\%\par
\begin{itemize}
\item Department A\FHPSpaces{2}(1) 
\item Department B\FHPSpaces{2}(2) 
\end{itemize}
\FHPBlockLabel{End of Block: Choice Experiment 3565\_8020}
\FHPBlockSeparator
\FHPBlockLabel{Start of Block: Choice Experiment 5050\_2080}
\par\noindent 5050\_2080 Which department do you believe would produce stronger faculty (produce better science)?\FHPLineBreak{}
\textbf{Weights in Department A:}\FHPLineBreak{}
Citation counts: 50\%\FHPLineBreak{}
Venue prestige: 50\%\FHPLineBreak{}
\textbf{Weights in Department B:}\FHPLineBreak{}
Citation count: 20\%\FHPLineBreak{}
Venue prestige: 80\%\par
\begin{itemize}
\item Department A\FHPSpaces{2}(1) 
\item Department B\FHPSpaces{2}(2) 
\end{itemize}
\FHPBlockLabel{End of Block: Choice Experiment 5050\_2080}
\FHPBlockSeparator
\FHPBlockLabel{Start of Block: Choice Experiment 5050\_8020}
\par\noindent 5050\_8020 Which department do you believe would produce stronger faculty (produce better science)?\FHPLineBreak{}
\textbf{Weights in Department A:}\FHPLineBreak{}
Citation counts: 50\%\FHPLineBreak{}
Venue prestige: 50\%\FHPLineBreak{}
\textbf{Weights in Department B:}\FHPLineBreak{}
Citation count: 80\%\FHPLineBreak{}
Venue prestige: 20\%\par
\begin{itemize}
\item Department A\FHPSpaces{2}(1) 
\item Department B\FHPSpaces{2}(2) 
\end{itemize}
\FHPBlockLabel{End of Block: Choice Experiment 5050\_8020}
\FHPBlockSeparator
\FHPBlockLabel{Start of Block: Choice Experiment 6535\_2080}
\par\noindent 6535\_2080 Which department do you believe would produce stronger faculty (produce better science)?\FHPLineBreak{}
\textbf{Weights in Department A:}\FHPLineBreak{}
Citation counts: 65\%\FHPLineBreak{}
Venue prestige: 35\%\FHPLineBreak{}
\textbf{Weights in Department B:}\FHPLineBreak{}
Citation count: 20\%\FHPLineBreak{}
Venue prestige: 80\%\par
\begin{itemize}
\item Department A\FHPSpaces{2}(1) 
\item Department B\FHPSpaces{2}(2) 
\end{itemize}
\FHPBlockLabel{End of Block: Choice Experiment 6535\_2080}
\FHPBlockSeparator
\FHPBlockLabel{Start of Block: Choice Experiment 6535\_5050}
\par\noindent 6535\_5050 Which department do you believe would produce stronger faculty (produce better science)?\FHPLineBreak{}
\textbf{Weights in Department A:}\FHPLineBreak{}
Citation counts: 65\%\FHPLineBreak{}
Venue prestige: 35\%\FHPLineBreak{}
\textbf{Weights in Department B:}\FHPLineBreak{}
Citation count: 50\%\FHPLineBreak{}
Venue prestige: 50\%\par
\begin{itemize}
\item Department A\FHPSpaces{2}(1) 
\item Department B\FHPSpaces{2}(2) 
\end{itemize}
\FHPBlockLabel{End of Block: Choice Experiment 6535\_5050}
\FHPBlockSeparator
\FHPBlockLabel{Start of Block: Choice Experiment 8020\_3565}
\par\noindent 8020\_3565 Which department do you believe would produce stronger faculty (produce better science)?\FHPLineBreak{}
\textbf{Weights in Department A:}\FHPLineBreak{}
Citation counts: 80\%\FHPLineBreak{}
Venue prestige: 20\%\FHPLineBreak{}
\textbf{Weights in Department B:}\FHPLineBreak{}
Citation count: 35\%\FHPLineBreak{}
Venue prestige: 65\%\par
\begin{itemize}
\item Department A\FHPSpaces{2}(1) 
\item Department B\FHPSpaces{2}(2) 
\end{itemize}
\FHPBlockLabel{End of Block: Choice Experiment 8020\_3565}
\FHPBlockSeparator
\FHPBlockLabel{Start of Block: Choice Experiment 8020\_6535}
\par\noindent 8020\_6535 Which department do you believe would produce stronger faculty (produce better science)?\FHPLineBreak{}
\textbf{Weights in Department A:}\FHPLineBreak{}
Citation counts: 80\%\FHPLineBreak{}
Venue prestige: 20\%\FHPLineBreak{}
\textbf{Weights in Department B:}\FHPLineBreak{}
Citation count: 65\%\FHPLineBreak{}
Venue prestige: 35\%\par
\begin{itemize}
\item Department A\FHPSpaces{2}(1) 
\item Department B\FHPSpaces{2}(2) 
\end{itemize}
\FHPBlockLabel{End of Block: Choice Experiment 8020\_6535}
\FHPBlockSeparator
\FHPBlockLabel{Start of Block: Self-Placement}
\par\noindent Q18 Which of the following hypothetical departments most resembles your department’s hiring practice?\par
\begin{itemize}
\item Citation count weight: \textbf{0\%} Venue quality weight: \textbf{100\%}\FHPSpaces{2}(1) 
\item Citation count weight: \textbf{10\%} Venue quality weight: \textbf{90\%}\FHPSpaces{2}(2) 
\item Citation count weight: \textbf{20\%} Venue quality weight: \textbf{80\%}\FHPSpaces{2}(3) 
\item Citation count weight: \textbf{30\%} Venue quality weight: \textbf{70\%}\FHPSpaces{2}(4) 
\item Citation count weight: \textbf{40\%} Venue quality weight: \textbf{60\%}\FHPSpaces{2}(5) 
\item Citation count weight: \textbf{50\%} Venue quality weight: \textbf{50\%}\FHPSpaces{2}(6) 
\item Citation count weight: \textbf{60\%} Venue quality weight: \textbf{40\%}\FHPSpaces{2}(7) 
\item Citation count weight: \textbf{70\%} Venue quality weight: \textbf{30\%}\FHPSpaces{2}(8) 
\item Citation count weight: \textbf{80\%} Venue quality weight: \textbf{20\%}\FHPSpaces{2}(9) 
\item Citation count weight: \textbf{90\%} Venue quality weight: \textbf{10\%}\FHPSpaces{2}(10) 
\item Citation count weight: \textbf{100\%} Venue quality weight: \textbf{0\%}\FHPSpaces{2}(11) 
\end{itemize}
\FHPBlockLabel{End of Block: Self-Placement}
\FHPBlockSeparator
\FHPBlockLabel{Start of Block: Block 14}
\par\noindent Q21 (Optional) Feel free to leave any thoughts you have about the survey or about venue prestige in general.\FHPLineBreak{}
Please reach out to nyuad.ai-and-society@nyu.edu if you'd like to get in touch.\par
\par\noindent \_\_\_\_\_\_\_\_\_\_\_\_\_\_\_\_\_\_\_\_\_\_\_\_\_\_\_\_\_\_\_\_\_\_\_\_\_\_\_\_\_\_\_\_\_\_\_\_\_\_\_\_\_\_\_\_\_\_\_\_\_\_\_\_\par
\par\noindent \_\_\_\_\_\_\_\_\_\_\_\_\_\_\_\_\_\_\_\_\_\_\_\_\_\_\_\_\_\_\_\_\_\_\_\_\_\_\_\_\_\_\_\_\_\_\_\_\_\_\_\_\_\_\_\_\_\_\_\_\_\_\_\_\par
\par\noindent \_\_\_\_\_\_\_\_\_\_\_\_\_\_\_\_\_\_\_\_\_\_\_\_\_\_\_\_\_\_\_\_\_\_\_\_\_\_\_\_\_\_\_\_\_\_\_\_\_\_\_\_\_\_\_\_\_\_\_\_\_\_\_\_\par
\par\noindent \_\_\_\_\_\_\_\_\_\_\_\_\_\_\_\_\_\_\_\_\_\_\_\_\_\_\_\_\_\_\_\_\_\_\_\_\_\_\_\_\_\_\_\_\_\_\_\_\_\_\_\_\_\_\_\_\_\_\_\_\_\_\_\_\par
\par\noindent \_\_\_\_\_\_\_\_\_\_\_\_\_\_\_\_\_\_\_\_\_\_\_\_\_\_\_\_\_\_\_\_\_\_\_\_\_\_\_\_\_\_\_\_\_\_\_\_\_\_\_\_\_\_\_\_\_\_\_\_\_\_\_\_\par
\FHPBlockLabel{End of Block: Block 14}
\FHPBlockSeparator
\endgroup

\clearpage
\newcommand{\FHPTitle}[1]{%
  \par\noindent{\fontsize{18}{22}\selectfont\bfseries #1\par}\vspace{1em}%
}

\FHPTitle{Extended Methods Section 3: Inferring respondent-level preferences from paired choices}

We inferred each respondent's underlying preference from their ten paired comparisons. Each comparison presented two hypothetical departments characterized by a citation-count weight, \(c\), and the complementary venue-prestige weight, \(100-c\). The offered citation weights were 20\%, 35\%, 50\%, 65\%, and 80\%.

We assumed that each respondent's choices could be represented by a single ideal citation weight, \(\theta_i\), and that the respondent would prefer the department whose citation weight was closer to that ideal. For comparison \(j\), let \(c_{jL}<c_{jH}\) denote the lower and higher citation weights, respectively. The model predicts selection of the higher citation weight when

\[
\theta_i > t_j=\frac{c_{jL}+c_{jH}}{2}.
\]

We coded \(y_{ij}=1\) when respondent \(i\) selected the option with the higher citation weight, regardless of whether that option appeared on the left or right. When \(\theta_i=t_j\), the two options are equally distant from the ideal; for coding purposes, we assigned the boundary to the lower interval.

The distinct comparison midpoints were 27.5\%, 35\%, 42.5\%, 50\%, 57.5\%, 65\%, and 72.5\%. The ten choices therefore distinguish eight ideal-weight intervals, expressed in percentage points:

\[
\begin{aligned}
\mathcal{I}_1 &= [0,27.5], &
\mathcal{I}_2 &= (27.5,35], &
\mathcal{I}_3 &= (35,42.5], &
\mathcal{I}_4 &= (42.5,50],\\
\mathcal{I}_5 &= (50,57.5], &
\mathcal{I}_6 &= (57.5,65], &
\mathcal{I}_7 &= (65,72.5], &
\mathcal{I}_8 &= (72.5,100].
\end{aligned}
\]

Every ideal weight within a given interval implies the same predicted response profile across the ten comparisons. Let \(p_{jk}\) denote the response predicted for comparison \(j\) under interval \(k\). For respondent \(i\) and interval \(k\), we calculated the number of choices that disagreed with this profile:

\[
L_{ik}=\sum_{j=1}^{10}\mathbf{1}\!\left(y_{ij}\neq p_{jk}\right).
\]

The respondent's minimum violation count was

\[
V_i=\min_k L_{ik},
\]

and the set of best-fitting intervals was

\[
B_i=\operatorname*{arg\,min}_k L_{ik}.
\]

Thus, \(V_i=0\) indicates that all ten choices were consistent with at least one ideal-weight interval, whereas \(V_i=1\) indicates that nine of the ten choices agreed with the closest-fitting profile. Of the 869 respondents, 603 (69.4\%) had \(V_i=0\), and 795 (91.5\%) had \(V_i\leq1\). Figure~\ref{fig}a uses all 869 respondents; analyses requiring a respondent-level ideal-weight interval use the 795 respondents with \(V_i\leq1\). This criterion permits one potentially noisy or inconsistent response while requiring the remaining choices to conform to a common underlying ideal.

When several intervals produced the same minimum violation count, we divided the respondent's contribution equally among them:

\[
w_{ik}=
\begin{cases}
1/\lvert B_i\rvert, & k\in B_i,\\
0, & k\notin B_i.
\end{cases}
\]

Every retained respondent therefore contributed a total weight of one. Let \(G_g\) denote the set of retained respondents in group \(g\), and let \(n_g=\lvert G_g\rvert\). The percentage assigned to interval \(k\) was

\[
S_{gk}=\frac{100}{n_g}\sum_{i\in G_g}w_{ik}.
\]

These percentages sum to 100\% within each group and form the distributions shown in Figure~\ref{fig}b--c.

The paired choices identify intervals rather than exact ideal weights. For descriptive point summaries, we represented the six interior intervals by their midpoints and bounded the two outer intervals at the lowest and highest offered citation weights, 20\% and 80\%. This produced the representative values

\[
m_k\in
\{23.75,31.25,38.75,46.25,53.75,61.25,68.75,76.25\}.
\]

For respondent \(i\), the choice-implied representative citation weight was

\[
\widetilde{\theta}_i=\sum_{k=1}^{8}w_{ik}m_k.
\]

Group means were obtained by averaging \(\widetilde{\theta}_i\) across retained respondents. Because the choices cannot distinguish among ideal weights within an interval, these means should be interpreted as midpoint-coded descriptive summaries rather than point-identified preferences.

Let \(q_i\) denote respondent \(i\)'s reported departmental citation weight. We defined the respondent-level signed preference--practice difference as

\[
d_i=\widetilde{\theta}_i-q_i,
\]

such that \(d_i>0\) indicates a preference for greater citation weight than the respondent attributed to their department. For group \(g\), the mean difference was

\[
\Delta_g=\frac{1}{n_g}\sum_{i\in G_g}d_i
        =\overline{\widetilde{\theta}}_g-\overline{q}_g.
\]

The paired-samples \(t\)-test reported in the Results tests whether the overall mean of \(d_i\) differs from zero.

\end{document}